\documentclass[%
 reprint,          
 aps,              
 prl,              
 superscriptaddress, 
 floatfix
]{revtex4-2}

\usepackage{graphicx}
\usepackage{dcolumn}
\usepackage{bm}
\usepackage{xspace}
\usepackage{multirow}
\usepackage{booktabs}
\usepackage{xcolor}
\usepackage{xfrac}

\usepackage{comment}
\usepackage{physics}
\usepackage{xcolor}

\usepackage[separate-uncertainty = true]{siunitx}
\DeclareSIUnit{\depth}{\gram\per\square\centi\meter}

\usepackage{hyperref}
\hypersetup{
    colorlinks=true,
    linkcolor=blue,
    filecolor=blue,      
    urlcolor=blue,
    pdftitle={Overleaf Example},
    pdfpagemode=FullScreen,
    }

\usepackage[mathlines]{lineno}

\newcommand{\suppref}{\hyperref[sec:SM]{Supplemental Material}}

\newcommand{\alphahad}{\alpha_{\text{had}}}

\newcommand{\zetahad}{\zeta_{\text{had}}}
\newcommand{\zetaem}{\zeta_{\text{EM}}}
\newcommand{\multhad}{m_{\text{had}}}
\newcommand{\multem}{m_{\text{EM}}}

\newcommand{\conex}{\texttt{Conex}}
\newcommand{\epos}{E\textsc{pos\,}LHC\xspace}
\newcommand{\qII}{QGS\textsc{jet}\,-II.04\xspace}

\newcommand{\eposr}{E\textsc{pos\,}LHC-R\xspace}
\newcommand{\qIII}{QGS\textsc{jet}\,-III.01\xspace}
\newcommand{\sibe}{S\textsc{ibyll}2.3e\xspace}

\newcommand{\xmax}{X_{\max}}

\newcommand{\nmu}{n_\mu}

\newcommand{\Nmu}{N_\mu}

\newcommand{\x}[1]{%
  {}$
  \kern-2\mathsurround 
  $
  \binoppenalty10000 \relpenalty10000 #1
  {}$
  \kern-2\mathsurround 
  $
}

\begin{document}


\title{
Forward hadron production in proton–air collisions above LHC energies through the fluctuations of extensive air showers
}

\author{Lorenzo Cazon}
\address{Instituto Galego de Física de Altas Enerxías (IGFAE),\\ Rúa de Xoaquín Díaz de Rábago, s/n, Campus Vida, Universidade de Santiago de Compostela, 15705, Santiago de Compostela, Galicia, Spain}

\author{Ruben Concei\c{c}\~{a}o}
\address{Departamento de F\'isica, Instituto Superior T\'ecnico (IST), Universidade de Lisboa, Av.\ Rovisco Pais 1, 1049-001 Lisbon, Portugal}
\address{Laborat\'{o}rio de Instrumenta\c{c}\~{a}o e F\'{i}sica Experimental de Part\'{i}culas (LIP) - Lisbon, Av.\ Prof.\ Gama Pinto 2, 1649-003 Lisbon, Portugal}

\author{Miguel A. Martins}
\address{Instituto Galego de Física de Altas Enerxías (IGFAE),\\ Rúa de Xoaquín Díaz de Rábago, s/n, Campus Vida, Universidade de Santiago de Compostela, 15705, Santiago de Compostela, Galicia, Spain}
\email{miguelalexandre.jesusdasilva@usc.es}

\author{Felix Riehn}
\address{Technische Universität Dortmund, August-Schmidt-Straße 4, 44221 Dortmund, Germany}

\date{\today}

\begin{abstract}

Primary proton--air interactions at ultra-high energies leave a physically interpretable imprint on the correlated fluctuations of the depth of shower maximum and the muon content in extensive air showers. This imprint reflects the stochasticity in the partition of the primary energy among secondary particles in the first interaction. We show that these fluctuations can be accessed through a probabilistic description that isolates sensitivity to hadronic physics in the initial collision, while treating the subsequent shower development as effectively universal. The uncertainties resulting from this universality are smaller than the spread among current hadronic interaction models and comparable to current experimental uncertainties. Consequently, the joint observable space defined by these two quantities provides a new probe of hadron production in kinematic regimes far beyond the reach of human-made accelerators.

\end{abstract}

\pacs{Valid PACS appear here}
\maketitle



\maketitle

\section{Introduction}
\label{sec:intro}
The origin and acceleration mechanisms of ultra-high-energy cosmic rays (UHECRs) remain a central open problem in astroparticle physics~\cite{2023_Snowmass_UHECR}. Due to their extremely low flux, UHECRs cannot be observed directly, and their properties must be inferred from the extensive air showers (EAS) they produce in the Earth’s atmosphere. These cascades, initiated at centre-of-mass energies exceeding $\SI{100}{\TeV}$, are governed by hadronic interactions that cannot be fully computed \textit{ab initio} from perturbative Quantum Chromodynamics. Consequently, interpreting air-shower data relies on phenomenological models that extrapolate accelerator measurements into poorly constrained kinematic regimes. The resulting model uncertainties prevent both a consistent~\cite{2022_Albrecht_MuonPuzzle, 2024_Auger_Xmaxs1000fits} and precise~\cite{2023_Thomas_XmaxFDrec} determination of the UHECR mass composition and hinder the identification of their sources. Progress, therefore, requires air-shower observables that retain direct and physically interpretable information on the mechanisms of particle production.

Two key observables routinely measured by modern air-shower experiments are the atmospheric depth of the shower maximum, $\xmax$, and the muon content. For fixed primary energy and mass composition, the stochastic natures of hadron production, reinteraction, and decay lead to intrinsic fluctuations in these observables. Traditionally, their distributions have been characterized through low-order moments, either to infer the primary mass~\cite{2012_Kampert_MassModelDep} or to constrain hadronic-interaction properties. These include the proton--air cross section~\cite{2012_Auger_xsection}, multiparticle production variables such as multiplicity and elasticity modified through \textit{ad hoc} resampling~\cite{2011_Ulrich_f19}, the fraction of energy transferred to neutral pions in the first interaction~\cite{2018_Cazon_alpha} and the hardness of their energy spectrum~\cite{2021_Cazon_lambdamu,2024_Martins_lambdamu_xmax}, and the cumulative nature of muon production from hadronic-cascade decays~\cite{2018_Cazon_alpha}.

The expected muon content reflects the cumulative contribution of many interactions throughout the shower development, while its stochastic fluctuations are dominated by variations in the energy partition in the first interaction~\cite{2018_Cazon_alpha}. In contrast, both the scale and fluctuations of $\xmax$ are primarily determined by the proton--air cross section and the secondary energy spectra of the first proton--air interaction~\cite{2016_Ostapchenko_XmaxScale_1stInt, 2025_miguel_xmaxmodel}. Separately, these features provide clues to resolve current tensions between air-shower data and hadronic-interaction model predictions~\cite{2022_Albrecht_MuonPuzzle, 2024_Auger_Xmaxs1000fits}. However, a more detailed description of hadronic interactions requires a connection between hadron production and the correlated fluctuations of $\xmax$ and the number of muons.

\begin{figure*}[!t]
    \centering
    \includegraphics[width=\linewidth]{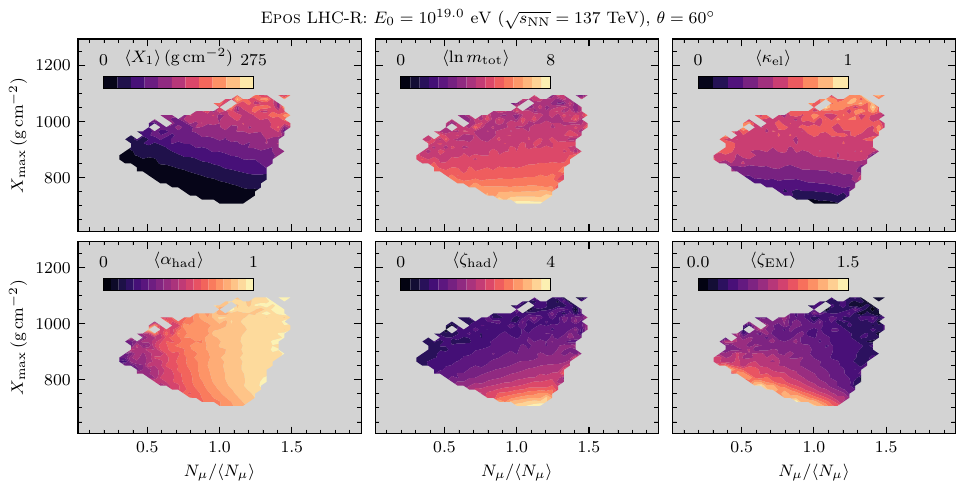}
    \caption{Mean values of several multiparticle--production variables of primary proton--air interactions across the joint distribution of $\nmu$ and $\xmax$. The top row shows classical variables, and the bottom row shows variables encoding information about the shape of the secondary--particle energy spectra. Produced with the ensemble of \conex{} simulations described in the Supplemental Material using \eposr{}.}
    \label{fig:primary_vars_nmuxmax_plane}
\end{figure*}


In this Letter, we show, for the first time, that the two-dimensional distribution of the depth of shower maximum and the number of muons is a powerful probe of forward hadron production in kinematic regimes inaccessible to current collider experiments. Its structure is largely determined by stochastic fluctuations in the energy distribution among secondaries of the primary interaction and in the partition of energy between its hadronic and electromagnetic channels. We show that the subdominant and weakly model-dependent fluctuations in the subsequent shower evolution can be treated as effectively universal, enabling a direct connection between physically consistent distributions of primary-interaction properties and the two-dimensional plane of air-shower observables. This universality is realized by replacing the explicit hadronic-interaction-model dependence with the average behavior of the first interaction, allowing quantitative discrimination among hadron-production mechanisms at energies above $100\,\mathrm{PeV}$ or $\sqrt{s}\gtrsim 14\,\mathrm{TeV}$. Finally, the uncertainty induced by this universality is below current experimental uncertainties~\cite{2021_Auger_muonfluctuations,2025_miguel_xmaxmodel}.

\section{Primary-interaction physics in the shower-observable plane}
\label{sec:first_int_in_nmuxmax_plane}

In proton--induced extensive air showers, the development of the cascade is largely determined by the properties of the first hadronic interaction. In particular, the partition of the primary energy between hadronic and electromagnetic channels, together with the energy spectrum of the secondary particles produced in that interaction, governs both the longitudinal development of the shower~\cite{2025_miguel_xmaxmodel} and the eventual muon content~\cite{2018_Cazon_alpha}. These considerations motivate the search for shower observables that retain sensitivity to the primary interaction on a shower--to--shower basis.

To characterize the primary proton--air interaction, we consider a set of variables that describe its point of interaction, multiplicity, and energy flow. Specifically, the distribution of the depths of first interaction points, $X_1$, reflects the proton--air interaction cross section. The total multiplicity of secondary particles produced in the interaction is denoted by $m_{\mathrm{tot}}$, while the elasticity, $\kappa_{\mathrm{el}}$, is defined as the fraction of the primary energy carried by the most energetic secondary particle.  

To quantify how energy is partitioned among secondaries, we separate the final state particles of the primary proton-air interaction into a hadronic sector, consisting of charged hadrons that predominantly re--interact (mostly charged pions, kaons and light baryons), and an electromagnetic sector dominated by the neutral mesons $\pi^0$ and $\eta$ and their decay products: $\gamma$ and $e^\pm$.
We denote the fraction of the primary energy $E_0$ carried by the $i$-th hadronically interacting secondary particle by $x_i \equiv E_i/E_0$ (and analogously $x_j$ for particles in the electromagnetic sector). The fraction of the primary energy transferred to the hadronic sector is denoted by $\alphahad \equiv \sum_{i=1}^{\multhad} x_i$~\cite{2018_Cazon_alpha}. In addition, we introduce production variables~\cite{2025_miguel_xmaxmodel} constructed from the energy fractions $x$ of secondary particles. For the hadronic sector, we define $\zetahad \equiv -\sum_{i=1}^{\multhad} x_i \ln x_i$, which is large when energy is evenly shared among many secondaries and small when a leading particle carries most of the primary energy, as in diffractive interactions. The analogous quantity for the electromagnetic sector is denoted by $\zetaem \equiv -\sum_{j=1}^{\multem} x_j \ln x_j,$. These production variables are sensitive to the shapes of the secondary energy spectra and were deduced to maximise the correlation between fluctuations in $\xmax$ and the stochastic production of hadrons in the primary interaction.

To suppress the dominant dependence of the muon content on the overall shower development, we introduce the relative muon content $\nmu \equiv \Nmu / \langle \Nmu \rangle$, where $\Nmu$ denotes the number of muons for each air-shower, and $\langle \Nmu \rangle$ its expected value over an ensemble of showers. 
The joint distribution of $\nmu$ and $\xmax$, denoted $f(\nmu,\xmax)$, provides a compact representation of the shower-to-shower fluctuations.

To make explicit how primary--interaction properties map onto this observable space, we examine the behavior of representative first--interaction variables across the $\nmu$--$\xmax$ plane.
Note that $\nmu$ is measured at ground and thus affected by the depth-dependent attenuation of the muon flux and its truncation by the ground. However, these propagation effects, which are modelled elsewhere~\cite{2012_Cazon_MuonTransport, 2025_Martins_PhDThesis}, do not mask the first-interaction imprint on the $n_\mu$--$X_{\max}$ correlation.

Figure~\ref{fig:primary_vars_nmuxmax_plane} shows mean values of the primary--interaction variables defined across the $\nmu$--$\xmax$ plane for simulated proton--induced air showers, with $E_0 = 10^{19}$ eV and zenith angle $\theta = 60^\circ$, using \conex{}~\cite{2004_Pierog_conex, 2007_Bergmann_conex} and the high-energy hadronic interaction model \eposr{}~\cite{2025_Tanguy_eposlhcr}. Further details of the simulation libraries are in the \suppref{}. Clear and approximately monotonic gradients are observed across the $\nmu-\xmax$ plane, demonstrating that the joint distribution of $\nmu$ and $\xmax$ retains information about the depth of the first interaction, the partition of the primary energy between the hadronic and electromagnetic sectors and the energy spectra of secondaries in each sector of the first interaction.

Shallow showers with enhanced relative muon content are associated with primary interactions in which a large fraction of the primary energy is transferred to the hadronic sector and evenly distributed among many secondary particles. This leads to a rapid dissipation of the primary energy and rapidly developing showers. These interactions are characterised by large values of $\zetahad$. Conversely, deep and muon--depleted showers arise from interactions in which a significant fraction of the energy is transferred to the electromagnetic sector, typically through energetic neutral pions. For these showers, the relative muon content is governed by $\alphahad$ while $\xmax$ is mostly determined by $\zetaem$. The deepest showers correspond to quasi--elastic interactions, in which a leading particle carries most of the primary energy, and re-interacts after an interaction length.

The structure of the $\nmu$--$\xmax$ plane therefore reflects well--defined classes of primary proton--air interactions. This establishes the joint distribution $f(\nmu,\xmax)$ as a physically interpretable observable space for probing hadron production in the first interaction using air--shower data.


\begin{figure*}[!t]
    \centering
    \includegraphics[width=\linewidth]{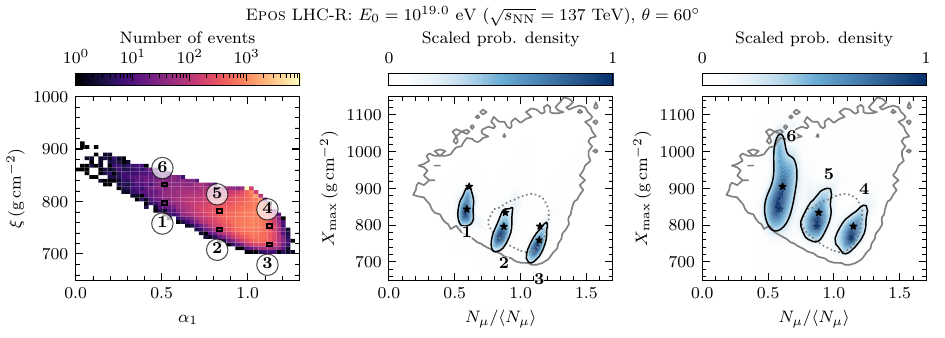}
    \caption{Left: joint distribution of $\alpha_1$ and $\xi$. Middle and right: examples of joint distributions of $\nmu$ and $\xmax$ corresponding to the particular values of $(\alpha_1,\xi)$ in the bins highlighted in the left panel. Star markers indicate the centroids; black contours show the $1\sigma$ boundary of each response; grey contours show the $1\sigma$ (dotted) and $2\sigma$ (solid) boundaries of the full $f(\nmu,\xmax)$.}
    \label{fig:shower_response_nmuxmax}
\end{figure*}

\section{Universality of the rest of the shower}
\label{sec:response}

The ordered structure of the gradients of production variables across the $\nmu-\xmax$ plane implies that information from the primary proton--air interaction survives subsequent shower evolution. To quantify this statement, we introduce two predictors of the shower development based on the spectra of hadrons of the primary interaction: a compound variable, $\xi$, highly correlated with $\xmax$, and a variable similar to $\alphahad$, $\alpha_1$, which correlates strongly with the relative muon content.

Following Ref.~\cite{2025_miguel_xmaxmodel}, the primary-interaction variable $\xi$ encompasses the fluctuations in the energy transferred to the hadronic and electromagnetic sectors, and in its partition among the secondaries of each sector, to estimate the value of $\xmax$ of the corresponding cascade. This variable has the form
\begin{equation}
\xi = \xi_0
      - A(\omega)\,\alpha_{\mathrm{had}}
      - \lambda_r\left(\omega\,\zeta_{\mathrm{had}} + \zeta_{\mathrm{EM}}\right),
\label{eq:xi_structural}
\end{equation}
where $\omega$ accounts for the average re-interaction of charged hadrons before producing the neutral pions that contribute to $\xmax$ and $\xi_0$ is the Heitler prediction for the $\xmax$ of a photonic cascade with energy $E_0 / 2$~\cite{1936_Heitler_Hmodel}. For the primary energy $E_0 = 10^{19.0}$ eV, we have $\xi_0 = \SI{917}{\depth}$,
$A = \left(311 - 9.6\,\omega\right)\unit{\depth}$ with $\omega = -0.140\,\langle \zeta_{\mathrm{EM}}/(1-\alpha_{\mathrm{had}})\rangle + 1.09$, and $\lambda_r = \SI{37}{\depth}$ the radiation length in air~\cite{2016_Gaisser_CRbook}.

To capture the fluctuations of the muon content stemming from the primary interaction, we define
\begin{equation}
\alpha_1 \equiv \sum_{i=1}^{m_{\rm had}} x_i^{\beta},
\label{eq:alpha1}
\end{equation}
where the sum runs over hadronic secondaries with energy fractions $x_i$ and $\beta\simeq0.93$~\cite{2018_Cazon_alpha} follows the Heitler--Matthews scaling~\cite{2005_Matthews_HMmodel}. Each term is the relative muon yield of hadron $i$. 

Figure~\ref{fig:shower_response_nmuxmax} shows that for identical primary interactions in the $\alpha_1-\xi$ space, the corresponding distributions of $\xmax$ and $\nmu$ is narrow compared to the full extent of $f(\nmu,\xmax)$. Moreover, the centroids of these distributions track the ordering induced by primary--interaction fluctuations. In fact, most of the dispersion is along the positive diagonal of the $\nmu-\xmax$ plane and results from fluctuations of $X_1$. These are uncorrelated with particle production in the primary interaction but correlated with the attenuation of the muon flux~\cite{2012_Cazon_MuonTransport, 2025_Martins_PhDThesis}.  Therefore, the anti-correlation between $\alpha_1$ and $\xi$, arising from the stochasticity of the energy transfer between the hadronic and EM sectors of the primary interaction, drives an anti-correlation between $\nmu$ and $\xmax$. These observations motivate a description of the shower, beyond the primary interaction, in which differences among hadronic-interaction models enter dominantly through the prior $f(\alpha_1, \xi)$. That is, in which the subsequent shower development acts as an additional universal smearing.

By modifying solely the prior of the primary interaction, $f(\alpha_1,\xi)$, through the choice of hadronic interaction model, we verified that for showers above $\SI{100}{\peta\eV}$ the universality of the subsequent shower development enables reproduction of the joint distribution $f(\nmu,\xmax)$ with enough accuracy to discriminate between different hadronic production mechanisms. This universality induces an intrinsic uncertainty in the leading moments of $f(\xmax)$ and $f(\nmu)$ amounting to $13\%$--$60\%$ of the spread of their predictions among hadronic interaction models. The most robustly predicted observables are $\expval{\xmax}$ and the steepness of the muon number distribution in muon-depleted showers, a mass-independent probe of the neutral-pion energy spectrum in the primary interaction~\cite{2021_Cazon_lambdamu}. Importantly, the uncertainty associated with the universality assumption is comparable to current experimental systematic uncertainties. These results are presented in detail in the \suppref{}. Consequently, the primary limitation for applying this framework to air-shower data is the present scarcity of ultra-high-energy events with high-quality reconstructions of both $\xmax$ and the muon content.

This Letter focuses on proton-induced air showers, for which the observed $\nmu$--$\xmax$ distribution is most directly tied to the properties of a single primary interaction. In data, however, the measured distribution is generally a superposition of primary masses. Nevertheless, the preservation of information about hadron production in the distribution of $\nmu$ and $\xmax$ for proton-induced showers justifies the exploration of air-shower data to exploit this deep connection, irrespective of the primary composition. Furthermore, the complexity of hadron production was successfully reduced to a simple description of a two-dimensional prior $f(\alpha_1, \xi)$. This reduction is likely generalizable to describe nuclear-air interactions. The applicability of the present findings to air-shower data can be envisioned in at least two complementary ways. First, the analysis can be applied to restricted event samples expected to be proton-dominated, such as deep and muon-depleted showers~\cite{2024_Martins_lambdamu_xmax}. Second, it can be combined with analyses that exhibit little or no sensitivity to the assumed primary-mass composition. For example, Ref.~\cite{2024_Auger_Xmaxs1000fits} showed that the additional deepening of $\xmax$ required to describe data from the Pierre Auger Observatory~\cite{2015_Auger_PAODescription} is largely independent of composition. In such cases, the findings introduced here can be used to measure the properties of the primary interaction.

\section{Conclusions}
\label{sec:conclusions}

We have shown that the joint shower-to-shower distribution of the relative muon content and the depth of shower maximum, $f(\nmu,\xmax)$, retains direct and physically interpretable information about multiparticle production in the primary proton--air interaction. The ordered structure of the $\nmu$--$\xmax$ plane encodes both the energy partition between hadronic and electromagnetic channels and features of the secondary--particle energy spectra.

By introducing the effects of the rest of the shower in terms of $(\alpha_1,\xi)$, we have demonstrated how physically consistent primary--interaction priors are mapped onto the observable $\nmu-\xmax$ plane with minimal explicit hadronic--model dependence. The intrinsic uncertainties of this mapping are small compared to the spread in current hadronic-interaction-model predictions and experimental systematics, preserving sensitivity to variations in hadron--production mechanisms at energies far beyond accelerator reach.

The results presented here are directly testable with current and upcoming data, and their reach will expand along two well-defined directions. First, while current measurements already achieve competitive resolutions in $\xmax$ and $\nmu$—approximately $\SI{15}{\depth}$~\cite{2023_Thomas_XmaxFDrec} and $\sim15\%$~\cite{2021_Auger_muonfluctuations} above $10^{18.5}$~eV for the Pierre Auger Observatory—larger multi-hybrid exposures will populate the $\nmu$–$\xmax$ plane with substantially increased statistics and improved control of independent systematics. AugerPrime~\cite{2019_Castellina_AugerPrime} provides a concrete path toward this regime.

In addition, although the mixed mass composition of the flux reduces sensitivity to hadronic production properties in inclusive samples, composition-enriched selections (e.g., deep, muon-depleted showers) can enhance the proton fraction and thereby sharpen the probe of the first hadronic interactions.

These results establish the $\nmu$--$\xmax$ plane as a powerful observable space for inferring hadronic interaction properties using air--shower measurements. As a result, statistically significant discrepancies between measured and predicted distributions of $\nmu$ and $\xmax$ translate directly into constraints on the underlying properties of the primary proton--air interaction.


\section*{Acknowledgments}
 The authors thank Bruce Dawson, Enrique Zas, Gonzalo Parente and Jaime Alvarez-Muñiz for carefully reading this manuscript. We extended our gratitude to other members of the Auger-IGFAE, Auger-LIP, and the Pierre Auger Collaboration for their valuable insights throughout the different stages of this work. The authors thank Ministerio de Ciencia e Innovaci\'on/Agencia Estatal de Investigaci\'on
(PID2022-140510NB-I00 and RYC2019-027017-I), Xunta de Galicia (CIGUS Network of Research Centers,
Consolidaci\'on 2021 GRC GI-2033, ED431C-2021/22 and ED431F-2022/15),
and the European Union (ERDF). This work has been partially funded by Fundação para a Ciência e Tecnologia, Portugal, under project \url{https://doi.org/10.54499/2024.06879.CERN}. MAM acknowledges that the project that gave rise to these results received the support of a fellowship from ``la Caixa” Foundation (ID 100010434). The fellowship code is LCF/BQ/DI21/11860033.
F.R. has received funding from the European Union’s Horizon 2020 research and innovation programme under the Marie Skłodowska-Curie grant agreement No. 101065027.

\bibliography{references}

\clearpage 
\appendix
\onecolumngrid 
\section*{Supplemental Material}\label{sec:SM}

This Supplemental Material provides validation and robustness studies supporting the physical interpretation presented in the main text. In particular, we carefully define the probabilistic mapping between primary-interaction variables and the observable $(\nmu, \xmax)$ plane, and specify the procedure to remove its explicit dependence on the hadronic interaction model. Moreover, we quantify intrinsic uncertainties of the universal probabilistic mapping or \textit{shower response function} as a function of the primary energy, using hadronic-interaction models as physically consistent priors for the primary interaction.

\subsection{Simulation setup and observable definitions} \label{subsec:simulation_description}

All results are based on ensembles of proton-induced extensive air showers generated with the \conex{} 2v7.80 simulation framework~\cite{2004_Pierog_conex, 2007_Bergmann_conex}. Libraries of $10^5$ air-showers were simulated at fixed primary energies, ranging from $10^{17}$ eV to $10^{19.5}$ eV in steps of $\Delta \log_{10} (E_0 / \unit{\eV}) = 0.5$, with fixed zenith angle $\theta = 60^\circ$. These simulations employ the high-energy hadronic-interaction models \eposr{}, \qIII{}~\cite{2024_Ostapchenko_qgsIII}, and \sibe{}~\cite{2020_Felix_sibyll23d}. Additional simulations with \epos{}~\cite{2015_Pierog_eposlhc} and \qII{}~\cite{2011_Ostapchenko_qgsjet}, using \conex{} 2v7.50, are used exclusively for validation and are not included in the construction of probabilistic model of the response function of the shower.

The transition energy between the full Monte Carlo treatment of interaction and the numerical computation of shower profiles using cascade equations was set to $E_{\rm{trans}} = 0.005 E_0$. This ensures an accurate computation of fluctuations of air-shower observables. The depth of shower maximum, $\xmax$, is obtained from a Gaisser--Hillas fit to the longitudinal profile of charged particles. The number of muons at ground, $\Nmu$, is evaluated at a slant depth corresponding to a zenith angle of $60^\circ$, and the relative muon content is defined as $\nmu \equiv \Nmu / \langle \Nmu \rangle$. This choice suppresses large shower-to-shower fluctuations associated with ground truncation effects. Moreover, $\Nmu$ can be related to the number of produced muons by considering the attenuation and decay of the muon flux~\cite{2012_Cazon_MuonTransport} and simple analytic models~\cite{2025_Martins_PhDThesis}. 

\subsection{Shower response function}

For fixed values of the primary-interaction predictors $(\alpha_1,\xi)$, the conditional distribution
$f(\nmu, \xmax \mid \alpha_1, \xi)$ is constructed from the corresponding subset of simulated showers. This distribution is the \textit{response function of the shower}, in terms of $\xmax$ and $\nmu$, to the primary interaction prior $f(\alpha_1, \xi)$. This response function, which depends on the hadronic interaction model $M$, $f(\nmu, \xmax\mid \alpha_1,\xi, M)$, encodes fluctuations from all interactions beyond the first and defines the integral transformation from priors on the primary interaction to the observable plane,
\begin{equation}
f(\nmu, \xmax)=\iint f(\nmu, \xmax\mid \alpha_1,\xi, M)\, f(\alpha_1,\xi)\, \dd{\alpha_1}\dd{\xi},
\end{equation}
which can be rewritten as
\begin{equation}
f(\nmu, \xmax)\equiv f(\nmu, \xmax\mid \alpha_1,\xi, M)\otimes f(\alpha_1,\xi).
\end{equation}
In practice, the response is narrow: for all but highly diffractive primary interactions, the region enclosed by the $1\sigma$ contour of
$f(\nmu,\xmax\mid \alpha_1,\xi,M)$ is smaller than that of the full $f(\nmu,\xmax)$, and distinct regions of $f(\alpha_1,\xi)$ map into largely disjoint regions of the $\nmu$--$\xmax$ plane. The limited overlap that appears along the positive diagonal is primarily induced by fluctuations in the depth of first interaction $X_1$, which are decoupled from particle production in the primary interaction but correlated with the depth-dependent attenuation of the muon flux. As a consequence, the anti-correlation between $\alpha_1$ and $\xi$ is largely preserved in $f(\nmu,\xmax)$.

Having established the sub-dominant role of the rest of the shower in determining the shape of $f(\nmu, \xmax)$, we can remove its explicit dependence on the hadronic-interaction model in two-steps. First, we eliminate direct model dependence by averaging response functions over the most up-to-date models: \eposr{}, \qIII{}, and \sibe{}. Second, we replace the residual dependence on the detailed shower development by its average dependence on the primary interaction, using the linear calibration between 
\begin{equation}
\expval{\xi} \equiv \iint  \xi f(\alpha_1,\xi) \dd{\alpha_1}\dd{\xi}\quad\textrm{and}\quad \expval{\xmax}\equiv \iint  \xmax f(\nmu,\xmax) \dd{\nmu} \dd{\xmax},
\end{equation}
\par\noindent 
introduced in Ref.~\cite{2025_miguel_xmaxmodel} of the main text. This motivates defining a universal shower response function
$f(\nmu,\xmax\mid \alpha_1,\xi,\expval{\xi})$ that provides an estimated distribution of $\nmu$ and $\xmax$ for any physically consistent prior $f(\alpha_1,\xi)$, :
\begin{equation} \label{eq:pred_nmu_xmax_univ_kernel}
\widehat{f}(\nmu, \xmax) = f(\nmu,\xmax\mid \alpha_1,\xi,\expval{\xi})\otimes f(\alpha_1, \xi).
\end{equation}
This construction captures the idea that consistent modifications of the primary interaction induce proportional changes in subsequent shower generations, while retaining sensitivity to the primary-interaction priors.
\par 
In practice, $f(\alpha_1, \xi\mid\nmu, \xi, \expval{\xi})$ is computed by finely binning the $(\alpha_1,\xi)$ and modelling, for each bin, the corresponding distribution of $\nmu$ and $\xmax$ using a two-dimensional Gaussian Kernel Density Estimation. The kernel bandwidth is computed using the Silverman rule~\cite{1986_Silverman_GKDE} and then multiplied by a factor to better reproduce the boundaries of $f(\nmu, \xmax)$, while avoiding artificial smoothing across low-density regions. Variations of the bandwidth within a reasonable range do not affect the conclusions drawn in the main text. 

\subsection{Validation of the universality of the shower response function}

To validate the probabilistic description, priors $f(\alpha_1,\xi)$ from individual hadronic-interaction models are forward-folded with the universal response function to obtain a predicted distribution $\widehat{f}(\nmu,\xmax)$ (see Equation~\eqref{eq:pred_nmu_xmax_univ_kernel}). This prediction is then compared to the Monte Carlo truth for the same model.

\begin{figure}[!ht]
    \centering
    \includegraphics[width=3.1in]{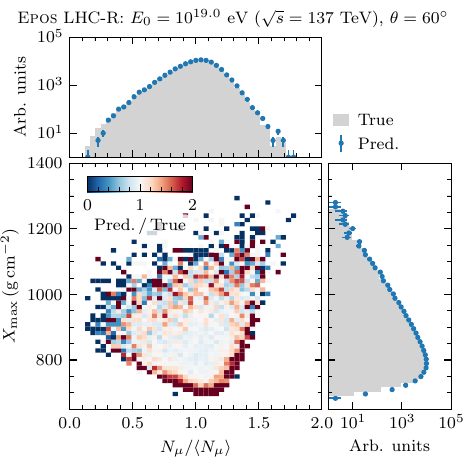}
    \caption{Central panel: ratio between the predicted and true $f(\nmu, \xmax)$. The prediction is obtained by applying Equation~\eqref{eq:pred_nmu_xmax_univ_kernel} to a prior $f(\alpha_1, \xi)$ provided by \eposr{}. Peripheral panels: True and predicted distributions of $\nmu$ (top) and $\xmax$ (right). The figure was produced using an ensemble of \conex{} simulations of proton-induced air showers with $E_0 = 10^{19}$ eV and $\theta = 60^\circ$, employing the high-energy hadronic interaction model \eposr{}.}
    \label{fig:pred_xmax_rmu_joint_dist}
\end{figure}

Figure~\ref{fig:pred_xmax_rmu_joint_dist} shows the ratio $\widehat{f}(\nmu,\xmax) / f(\nmu,\xmax)$ and the marginal true and predicted distributions of $\nmu$ and $\xmax$, for \eposr{}. The joint distribution and its marginals are reproduced with good accuracy. Deviations are confined to low-statistics regions at the edges of the distribution and are dominated by finite-sample effects and the modelling of the shower response using a kernel with finite width.

\subsection{Intrinsic uncertainties induced by the universal response function}

The intrinsic uncertainty of the universal response function is estimated by comparing the biases induced on selected moments of $f(\nmu)$ and $f(\xmax)$ when forward-folding priors from different hadronic-interaction models. Half of the spread of these biases is the inherent uncertainty of the model defined in Equation~\ref{eq:pred_nmu_xmax_univ_kernel}. The considered models were \eposr{}, \qIII{}, \sibe{}, and, importantly, \epos{} and \qII{}, which have not been used in the modelling of the universal shower response function.

We consider the mean and one-sided standard deviations of $\xmax$, the one-sided standard deviation of $\nmu$, and slopes of the exponential tails of $f(\xmax)$ (deep-tail with slope $\Lambda_\eta$~\cite{2012_Auger_xsection}) and $f(\nmu)$ (muon-depleted tail with slope $\Lambda_\mu$~\cite{2021_Cazon_lambdamu}).

\begin{figure}[ht!]
    \centering
    \includegraphics[width=\linewidth]{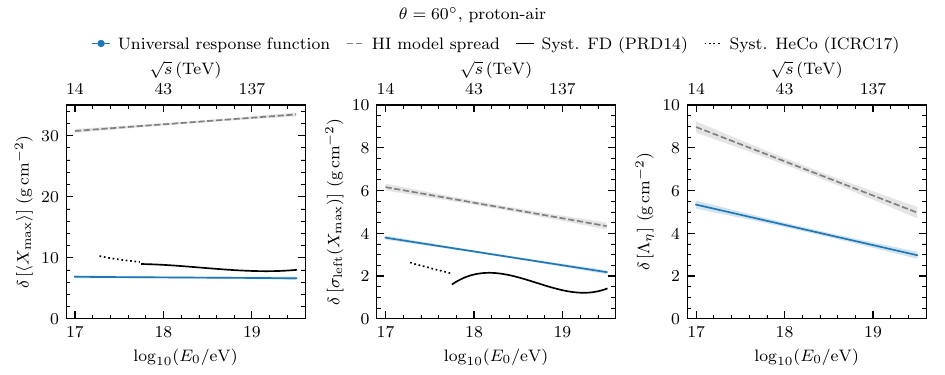}
    \caption{Primary energy evolution of the uncertainty on the main moments of $f(\xmax)$: $\delta\left[\expval{\xmax} \right]$ (left), $\delta \left[ \sigma_{\rm{left}}(\xmax)\right]$ (middle) and $\delta\left[\Lambda_\eta\right]$ (right). This uncertainty is represented in blue and it is half the spread of the biases induced on these moments by predicting $f(\nmu, \xmax)$ using Equation~\eqref{eq:pred_nmu_xmax_univ_kernel} on $f(\alpha_1, \xi)$ priors provided by: \eposr{}, \epos{}, \qIII{}, \qII{} and \sibe{}. All uncertainties are fitted to linear functions by minimizing $\chi^2$. The spread of model predictions for each moment isrepresented by the grey dashed line. The experimental systematic uncertainties are taken from Ref.~\cite{2018_Bellido_HeCoXmax} (``Syst. HeCo (ICRC17)") and Refs.~\cite{2014_Auger_xmax, 2023_Thomas_XmaxFDrec} (``Syst. FD (PRD14)"), and are represented in black.}
    \label{fig:xmax_pred_uncertainty_moments_penergy}
\end{figure}

\begin{figure}[ht!]
    \centering
    \includegraphics[width=5.5in]{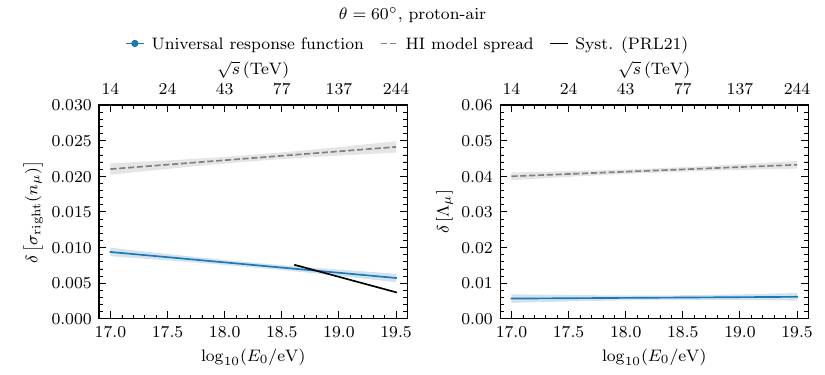}
    \caption{Primary energy evolution of the uncertainty on the main moments of $f(\nmu)$: $\delta \left[ \sigma_{\rm{right}}(\nmu)\right]$ (left) and $\delta\left[\Lambda_\mu\right]$ (right). This uncertainty is represented in blue and it is half the spread of the biases induced on these moments by predicting $f(\nmu, \xmax)$ using Equation~\eqref{eq:pred_nmu_xmax_univ_kernel} on $f(\alpha_1, \xi)$ priors provided by: \eposr{}, \epos{}, \qIII{}, \qII{} and \sibe{}. All uncertainties are fitted to linear functions by minimizing $\chi^2$. The spread of model predictions for each moment isrepresented by the grey dashed line. The experimental systematic uncertainty is taken from Ref.~\cite{2021_Auger_muonfluctuations} (``Syst. (PRL21)"), and it is represented in black.}
    \label{fig:rmu_pred_uncertainty_moments_penergy}
\end{figure}

Figures~\ref{fig:xmax_pred_uncertainty_moments_penergy} and \ref{fig:rmu_pred_uncertainty_moments_penergy} show the primary-energy dependence of the uncertainties on the lead moments of $\xmax$ and $\nmu$, respectively, over the range $10^{17}$--$10^{19.5}\,\mathrm{eV}$. The figures also show the spread of hadronic-interaction-model predictions of each model (depicted in grey). When available, the figures also display the experimental systematic uncertainties, using the case of the Pierre Auger Observatory~\cite{2015_Auger_PAODescription}. The systematic uncertainties on the scale and standard deviation of $\xmax$ using Fluorescence Detectors are taken from Refs.~\cite{2018_Bellido_HeCoXmax} and \cite{2014_Auger_xmax, 2023_Thomas_XmaxFDrec}. The systematic uncertainty on the standard deviation of $\nmu$ is taken from Ref.~\cite{2021_Auger_muonfluctuations}. Other observables either lack experimental measurements or these are being updated~\cite{2023_Olena_xsection_comp}. From an experimental perspective, these uncertainties define a benchmark precision in air-shower measurements at ultra-high energies.   
\par 

For all energies considered, the intrinsic uncertainty remains significantly smaller than the spread among current model predictions and is comparable to present experimental systematics. This ensures that the sensitivity to variations in hadron-production mechanisms of the primary interaction can be captured within the assumption of the universality of the rest of the shower. 

Overall, the larger uncertainties in $\sigma(\xmax)$ and $\Lambda_\eta$ arise from a combination of finite statistics in sparsely populated regions and model-dependent fluctuations in the depth of the first interaction, as well as in the interaction depth of the leading particle in diffractive first interactions. These fluctuations are largely uncorrelated with particle production in the primary interaction and are therefore not captured, by construction, by the universal response function. Moreover, as the primary energy decreases, these fluctuations become increasingly comparable to those of the primary interaction itself, explaining the growth of the uncertainty with decreasing $E_0$. These effects do not impact the main physical conclusions of the Letter.

\subsection{Summary}

The studies presented here demonstrate that the probabilistic description used in the main text provides a faithful and robust mapping between primary-interaction properties and the observable $(\nmu, \xmax)$ plane. The intrinsic uncertainties of the response are small compared to current model spreads and experimental systematics, supporting the interpretation of the structured geometry of $f(\nmu, \xmax)$ as a direct imprint of hadron production in the primary proton--air interaction.

\end{document}